\documentclass[12pt]{article}
\usepackage{a4,epsfig,multirow}

\newcommand{\Table}[1]{Table~\ref{#1}}
\newcommand{\Figure}[1]{Fig.\ref{#1}}
\newcommand{\diagram}[3]{\raisebox{-#3pt}{\epsfig{figure=#1,width=#2pt}}}
\newcommand{\lb}{\linebreak[1]}
\newcommand{\eee}[2]{$#1\times10^{#2}$}
\newcommand{\ie}{{\it i.e.\/}}
\newcommand{\eg}{{\it e.g.\/}}
\newcommand{\address}[1]{\\{\normalsize #1}}
\begin{document}
\title{Challenges in calculations for multi-particle processes%
\thanks{Presented at the final meeting of the European Network
        ``Physics at Colliders'', Mont\-pellier, September 26-27, 2004.}%
}
\author{A.\ van Hameren
\address{Institut f\"ur Physik, Johannes-Gutenberg-Universit\"at,
         Mainz, Germany}
\and
C.G.\ Papadopoulos
\address{Institute of Nuclear Physics, NCSR ``Demokritos'',
         15310 Athens, Greece}
}
\maketitle
\begin{abstract}
The physics of high-energy collider experiments asks for delicate
comparisons between theoretical predictions and experimental data.
Signals and potential backgrounds for new physics have to be predicted
at sufficient accuracy.  The accuracy as well as the computational
complexity of the calculations leading to the predictions depend on both
the number of external particles in the process analyzed and the order
of the quantum corrections, the number of loops, included in the
calculation. We present some approaches to problems occurring in these
calculations regarding the integration of phase-space and the inclusion
of one-loop corrections.
\end{abstract}

\section{Introduction}
%
Collider experiments played and continue to play a fundamental r\^ole in
particle physics.
This is exemplified by past, current and future experiments like, \eg\ LEP at
CERN, Tevatron at Fermilab or HERA at DESY, and LHC, again at CERN,
respectively.
Current research focuses on even more precise tests of what is known as the
Standard Model of particle physics, the validity of which needs the discovery
of the Higgs boson at such a collider experiment.
This asks for new colliders to reach higher energies leading to events from
scattering experiments with higher numbers of particles involved.

In order to prepare and analyze the outcome of the experiments, signals and
potential backgrounds for new physics have to be predicted at sufficient
accuracy.
Most of the calculations involved are set up within the framework of
perturbation theory, in which the accuracy is controlled by the order
parameter.
In the application to quantum field theory, on which the Standard Model is
based, the order parameter is connected to both the number of external
particles in the process analyzed and the order of the quantum corrections, the
number of loops, included in the calculation.

The Standard Model is tested mainly by the comparison of the experimental data
and theoretical calculations on the statistical level.
This can happen through the analysis of cross sections, or simulation of
collider experiments.
Both the increase in the multiplicity of particles at the experiments, and the
need for higher accuracy of theoretical calculations lead to a dramatic
increase of the complexity of this analysis.
In the following, we will encounter two particular examples in which this
increase of complexity constitutes a challenge in the scientific process.
%

\section{A hierarchical phase space generator for QCD antennas}
%
The reliable description of multi-jet production at collider experiments is an
important issue in the study of the Standard Model.
It requires the calculation of cross sections, which again requires the
integration of squared scattering amplitudes over phase space.
The number of dimensions of the integration space in combination with the
desired cuts point at the Monte Carlo method as the only suitable candidate for
this task.
Since the computation of QCD scattering matrix elements with many particles is
rather time-consuming, the integration process should preferably involve as few
integration points as possible.
The strong peaking structures exhibited in the QCD amplitudes enforce the
application of the Monte Carlo method to be dressed with a sophisticated
portion of importance sampling.
Flat phase space generators, like {\tt RAMBO}~\cite{rambo}, will not be
adequate for this task.

In the last years several methods to efficiently integrate the peaking
structures of the scattering amplitudes have emerged, and have been used in
several contexts~\cite{LEP2}.
For instance, {\tt PHEGAS}~\cite{phegas} is an example where an efficient,
automated, mapping of all possible peaking structures of a given scattering
process has been established.
The algorithm is based on the ``natural'' mappings dictated by the Feynman
graphs contributing to the given process, so that the number of kinematic
channels used to generate the phase space is equal to the number of Feynman
graphs.
Using adaptive methods, like multi-channel optimization~\cite{KleissPittau} and
by throwing away channels that are negligible, we may end up with a few channel
generator exhibiting high efficiency, as is indeed the case in
$n(+\gamma)$-fermion production in $e^+e^-$ collisions.
In contrast, the QCD scattering amplitudes point towards the opposite
direction: large number of Feynman graphs which means large number of
kinematic channels which, moreover, contribute equally to the result.

A way out off this problem may be based on the long-standing remark that
$n+2$-gluon amplitude may be described by a very compact expression when
special helicities are assigned to the gluons, which, combined with the leading
color approximation, results to
\begin{equation}
\sum_{c}|\mathcal{M}|^2
=8\left(\frac{N_{\mathrm{c}}}{2}\right)^{n} (N_{\mathrm{c}}^2-1)
\sum_{1\le i<j}^{n+2} (p_i\cdot p_j)^4
\!\!\!\!\sum_{P(2,\ldots,n+2)}\!\!\!\!A_{n+2}(p_1,\ldots,p_{n+2})
\;\;,
\end{equation}
where $N_{\mathrm{c}}$ refers to the number of colors,
\begin{equation}
A_{n+2}(p_1,\ldots,p_{n+2})
  =
[\;(p_1\cdot p_2)(p_2\cdot p_3)
         \cdots(p_{n+1}\cdot p_{n+2})(p_{n+2}\cdot p_1)\;]^{-1}
\;\;,
\label{defAnt}\end{equation}
and the sum over all permutations of the $2^{\mathrm{nd}}$ to the
$(n+2)^{\mathrm{nd}}$ argument of this function is taken, with the exception of
those that are equivalent under reflection $i\mapsto n+4-i$ \cite{kuijf}.

{\tt SARGE} \cite{sarge} is the first known example of a phase space generator
that deals with the momentum structures entering the above expression, namely
with (\ref{defAnt}), known as {\it antenna structures}.
The algorithm is based on the ``democratic'' strategy to generate the $n$ body
phase space, as is the case for {\tt RAMBO}, and it makes use of the scale
symmetry of the antenna to achieve the required goal.
Now, we study the ``hierarchical'' strategy for phase space
generation in order to efficiently map the momentum antenna structures.
The idea is as follows.
Using the standard two-body phase space (neglecting factors of $2\pi$)
\begin{equation}
d\Phi_2(P;s_1,s_2;p_1,p_2)
= d^4p_1\,\delta_+(p_1^2-s_1)\,d^4p_2\,\delta_+(p_2^2-s_2)\,\delta^4(P-p_1-p_2)
\;\;,
\label{twobody}\end{equation}
we decompose the phase space
\begin{equation}
d\Phi_n(P;p_1\ldots,p_n)
=\,\delta^4\Big(\sum_{i=1}^{n} p_i-P\Big)
     \prod_{i=1}^n d^4p_i\,\delta_+(p_i^2-\sigma_i)
\nonumber\end{equation}
as
\begin{eqnarray}
d\Phi_n(P;p_1\ldots,p_n)
&=&\,ds_{n-1}\,d\Phi_2(Q_n;\sigma_{n},s_{n-1};p_n,Q_{n-1})
\nonumber\\
&\times&\,ds_{n-2}\,d\Phi_2(Q_{n-1};\sigma_{n-1},s_{n-2};p_{n-1},Q_{n-2})
\nonumber\\
&&\hspace{20pt}\vdots\nonumber\\
&\times&\,ds_2\,d\Phi_2(Q_3;\sigma_3,s_2;p_3,Q_2)
\nonumber\\
&\times&\,d\Phi_2(Q_2;\sigma_2,\sigma_1;p_2,p_1)
\;\;.
\label{psdec}\end{eqnarray}
The task is to express the phase space in terms of the invariants $p_i\cdot
p_j$ appearing in the antenna structure (\ref{defAnt}), so that, using a
suitable mapping, we can construct a density that, apart from constant and soft
terms, will be identical to this antenna structure.
%

\subsection{Antenna generation}
%
In this section, we will present a sketch of how the antenna structures can be
generated using the hierarchical approach.
For a treatment in full detail, we refer to \cite{haag}.
The starting point is the generation of he two-body phase space (\ref{twobody})
in terms of the variables
\begin{equation}
a_1=\frac{q_1\cdot p_1}{q_1\cdot P}\quad,\quad
a_2=\frac{p_2\cdot q_2}{P\cdot q_2} \;\;,
\label{a1a2}
\end{equation}
where $q_1,q_2$ are given massless momenta.
Let us introduce the notation
\begin{equation}
\mathrm{c}
=\cos(\angle(\hspace{-1pt}\vec{\hspace{1pt}q}_1
            ,\hspace{-1pt}\vec{\hspace{1pt}q}_2))
\quad,\quad
\mathrm{s}=\sqrt{1-\mathrm{c}^2}
\end{equation}
and
\begin{equation}
s = P^2 \quad,\quad \bar{s}_{1,2}=s_{1,2}/s \;\;.
\end{equation}
We find that the parameterization
\begin{eqnarray}
   p_1^0 &\leftarrow& (s + s_1 - s_2)/(2\sqrt{s})
   \quad,\nonumber\\
   p_1^3 &\leftarrow& p_1^0 - \sqrt{s}\,a_1
   \quad,\nonumber\\
   p_1^2 &\leftarrow& (\;(\sqrt{s}-p_1^0-\sqrt{s}\,a_2)+\mathrm{c}p_1^3\;)/\mathrm{s}
   \quad,\\
   p_1^1 &\leftarrow& \pm(\;(p_1^0)^2-s_1-(p_1^2)^2-(p_1^3)^2\;)^{1/2}
   \quad,\nonumber
\end{eqnarray}
leads to the identity
\begin{equation}
d\Phi_2(P;s_1,s_2;p_1,p_2)=
da_1\,da_2\, \Pi(a_1,a_2)^{-1/2}\,\Theta(\,\Pi(a_1,a_2)\,)  \;\;,
\label{2bps}
\end{equation}
where
\begin{eqnarray}
\Pi(a_1,a_2) &=&
   4\mathrm{s}^2[(1-a_2+\bar{s}_2-\bar{s}_1)a_2-\bar{s}_2] \nonumber\\
   &-&[(1-2a_1-\bar{s}_1+\bar{s}_2)+(1-2a_2-\bar{s}_1+\bar{s}_2)\mathrm{c}]^2
   \;\;,
\label{defPi}\end{eqnarray}
and $\Theta$ is the step function.
So in order to obtain a two-body phase space with a density which depends on
the invariants $a_1,a_2$ following some given function $f(a_1,a_2)$, one has to
generate $a_1,a_2$ following a density proportional to
$f(a_1,a_2)\times\Pi(a_1,a_2)^{-1/2}$ in the region where $\Pi(a_1,a_2)>0$, and
construct the momenta as given above.

The generation strategy proceeds through a sequence of two-body phase space
generations following the decomposition (\ref{psdec}).
At each two-body generation, one final-state momentum $p_{k}$ is generated,
together with the sum $Q_{k-1}$ of the remaining final-state momenta to be
generated.
This suggest to label the momenta in a way opposite to the order of generation,
so first $p_{n},Q_{n-1}$ are generated, then $p_{n-1},Q_{n-2}$ and so on.
The starting point is the center of mass frame (CMF) of the initial momenta
$q_1$ and $q_2$ with $Q_n=q_1+q_2$ being the overall momentum.
The CMF of momentum $Q_k$ we denote by CMF$_k$.
The pair $p_{k},Q_{k-1}$ is generated by generating variables
$a_1^{(k)},a_2^{(k)}$ and constructing the momenta as described before.
These variables are now equal to
\begin{equation}
a_1^{(k)}=\frac{p_{k+1}\cdot p_k}{p_{k+1}\cdot Q_k}
\quad
\mathrm{and}
\quad
a_2^{(k)}=\frac{q_2\cdot Q_{k-1}}{q_2\cdot Q_k} \;\;.
\nonumber\end{equation}
This happens in CMF$_k$, so in order to obtain $p_{k},Q_{k-1}$, the constructed
momenta have to be boosted such that $(\sqrt{Q_k^2},0,0,0)$ is transformed to
$Q_k$.
Now, we do three observations.
Firstly, we have
\begin{equation}
  p_{k+1}\cdot Q_k=(Q^2_{k+1}-Q^2_{k}-p^2_{k+1})/2
\;\;.
\nonumber\end{equation}
Secondly, we have, with $\Sigma_{k}=\sum_{i=1}^{k}\sigma_{i}$, 
\begin{equation}
 \frac{s_{k}-\Sigma_{k}}
    {\left(s_{k}-\sigma_{k}-s_{k-1}\right)\left(s_{k-1}-\Sigma_{k-1}\right)}
 =  \frac{d}{ds_{k-1}}\,
    \log\left(\frac{s_{k-1}-\Sigma_{k-1}}{s_{k}-\sigma_{k}-s_{k-1}}\right)
  \;\;,
\label{s1g}
\end{equation}
and thirdly, we can write
\begin{eqnarray}
A_{n+2}(q_1,p_n,p_{n-1}\ldots,p_1,q_2)
&=& \frac{1}{2^{n-1}}(s_n-\Sigma_n)(q_1\cdot Q_n)(q_2\cdot Q_n)
\nonumber\\
&\times&\left(\prod_{k=n}^{3}g_{k}(s_{k-1})
        \frac{1}{a_1^{(k)}a_2^{(k)}}\right)\frac{1}{a_1^{(2)}a_2^{(2)}}  \;\;,
\label{int}
\end{eqnarray}
where $g_{k}(s_{k-1})$ is given by (\ref{s1g}), and where $s_{n}=Q^2_{n}$,
$p_{n+1}=q_1$ and $Q_1=p_1$.
These observations suggest that the phase space generation
\begin{eqnarray}
&& ds_{n-1}\, g_{n}(s_{n-1}) \;\;
   da^{(n)}_1\, \frac{1}{a^{(n)}_1}\;\;
   da^{(n)}_2\, \frac{1}{a^{(n)}_2}\;\;
   \Pi_{(n)}^{-1/2} \Theta (\Pi_{(n)})
\nonumber\\
&& ds_{n-2}\, g_{n-1}(s_{n-2})\;\;
   da^{(n-1)}_1\, \frac{1}{a^{(n-1)}_1}\;\;
   da^{(n-1)}_2\, \frac{1}{a^{(n-1)}_2}\;\;
   \Pi_{(n-1)}^{-1/2} \Theta (\Pi_{(n-1)})
\nonumber\\
&&\vdots
\nonumber\\
&& ds_2\, g_{3}(s_2)\;\;
   da^{(3)}_1 \, \frac{1}{a^{(3)}_1}\;\;
   da^{(3)}_2 \, \frac{1}{a^{(3)}_2}\;\;
   \Pi_{(3)}^{-1/2} \Theta (\Pi_{(3)})
\nonumber\\
&& da^{(2)}_1 \, \frac{1}{a^{(2)}_1}\;\;
   da^{(2)}_2 \, \frac{1}{a^{(2)}_2}\;\;
   \Pi_{(2)}^{-1/2} \Theta (\Pi_{(2)}) \;\;,
\label{closedA}
\end{eqnarray}
will lead to a density for the momenta that is proportional to $A_{n+2}$.
Three variables are generated in each CMF$_k$, namely $s_{k-1}$, $a^{(k)}_1$
and $a^{(k)}_2$.
Just as the integration of $s_{k-1}$ (\ref{s1g}), also the integration of
$a_1^{(k)},a_2^{(k)}$ results in a volume factor that depends on the
corresponding variables generated in CMF$_{k+1}$.
However, these factors are logarithmic functions of their arguments and exhibit
a non-singular behavior, and we call them {\em soft\/} factors.
The total actual density will therefore be the product of $n-1$ soft factors
times the antenna structure under consideration.

In the end, we want to generate all permutations in the momenta of
(\ref{int}).
Those for which $q_1$ and $q_2$ each appear in two factors (none of which is
$q_1\cdot q_2$) cannot be obtained by simple re-labeling.
In order to obtain these, we observe that they can be decomposed into two
antennas, namely
\begin{equation}
        A_{m+2}(q_1,p_{m},\ldots,p_{1},q_2)
  \times A_{n-m+2}(q_2,p_{n},\ldots,p_{m+1},q_1)
\label{otherA}
\end{equation}
and each of these can be generated after the decomposition,
\begin{eqnarray}
d\Phi_n(P;p_1\ldots,p_n)&=&ds_m\, ds_{n-m}\,
                           d\Phi_2(Q_n;s_{m},s_{n-m};Q_m,Q_{n-m})
\label{split}\\
&\times&d\Phi_m(Q_m;p_1,\ldots,p_m)\,d\Phi_{n-m}(Q_{n-m};p_{m+1},\ldots,p_n)
\;\;.
\nonumber
\end{eqnarray}
In order to combine the two sub-antennas to the required antenna structure, we
have to take into account in the first decomposition a density that is
proportional to
\begin{equation}
\frac{\Theta(\,\sqrt{s_{n}}-\sqrt{s_m}-\sqrt{s_{n-m}}\;)}{
(q_1\cdot Q_m)
(q_1\cdot Q_{n-m})
(q_2\cdot Q_m)
(q_2\cdot Q_{n-m})
\,
s_m
\,
s_{n-m}} \;\;.
\label{splitden}
\end{equation}
%

\subsection{Results}
%
In this section, we present results obtained by {\tt SARGE} and {\tt
HAAG}\footnote{{\tt HAAG} stands for: {\tt Hierarchical AntennA Generation}.},
the program that implements the hierarchical algorithm described before.
In order to be as general as possible, the only cut we apply is
\begin{equation}
   (p_i+p_j)^2 \ge s_0\;\;,
\nonumber\end{equation}
where $i,j(i\neq j)$ runs from $1$ to $n+2$ where $n$ is the number of
final-state particles.
Unless explicitly mentioned differently, we use $s_0=900\,\mathrm{GeV}^2$ and
the total energy $\sqrt{s}=1000\,\mathrm{GeV}$.
Moreover, all particles are assumed to be massless in order to compare with
{\tt SARGE}, with which only massless particles can be treated.

As it was mentioned in the introduction, we are interested in integrating sums
of QCD antenna structures (\ref{defAnt}).
We start by considering the simplest case, namely integrating the function
\begin{equation}
 s^2
 [\;(p_1\cdot p_3)(p_3\cdot p_4)(p_4\cdot p_2)(p_2\cdot p_5)
        \ldots(p_{n+2}\cdot p_1)\;]^{-1}
\label{ss3}
\end{equation}
that corresponds to a given permutation of the momenta, namely $(1,3,4,2,5,$
$\ldots,n+2)$.
In \Table{ss3t} we give the results for {\tt SARGE}, and {\tt HAAG}.
In all three codes the same single channel, corresponding to (\ref{ss3}), has
been used in the generation.
$N_{\mathrm{gen}}$ and $N_{\mathrm{acc}}$ are the number of generated and
accepted events, and by $f$
we define
\begin{equation}
  f = \frac{V_2}{I^2} \;\;,
\nonumber\end{equation}
where $V_2$ is the quadratic variance and $I$ is the estimated integral.
$f$ is clearly a measure of the efficiency of the generator.
Moreover $\varepsilon$, defined as
\begin{equation}
  \varepsilon=\frac{<w>}{w_{\mathrm{max}}} \;\;,
\nonumber\end{equation}
is the usual generation efficiency related for instance to `unweighted' events
in a realistic simulation.
\begin{table}[ht]
{\scriptsize
\begin{center}
\begin{tabular}{|c|c|c|c|c|c|c|c|}
\hline
jets & algorithm  & $N_{\mathrm{gen}}$  &  $N_{\mathrm{acc}}$  &  $I$         &   $\Delta I$  & $f$  &          $\varepsilon$(\%)  \\ \hline
\multirow{3}{1cm}{\centering 4}
    & {\tt SARGE}   &  \eee{1}{5}  & 34853   & \eee{.251}{-9} & \eee{.734}{-11} &     $85.9 $        & $0.34$ \\
    & {\tt HAAG}    &  \eee{5}{4}  & 31193   & \eee{.260}{-9} & \eee{.280}{-11} &     $5.75 $        & $1.77$   \\\hline
\multirow{3}{1cm}{\centering 5}
    & {\tt SARGE}   &  \eee{2.5}{5}  & 30960   & \eee{.438}{-10} & \eee{.153}{-11} &     $307  $        & $0.23$ \\
    & {\tt HAAG}    &  \eee{6.5}{4}  & 29855   & \eee{.442}{-10} & \eee{.640}{-12} &     $13.6 $        & $1.02$   \\\hline
\multirow{3}{1cm}{\centering 6}
    & {\tt SARGE}   &  \eee{1}{6}    & 28383   & \eee{.487}{-11} & \eee{.164}{-12} &     $1141 $        & $0.21$ \\
    & {\tt HAAG}    &  \eee{1.2}{5}  & 32070   & \eee{.487}{-11} & \eee{.658}{-13} &     $21.9 $        & $1.48$   \\\hline
\end{tabular}
\caption[.]{Results for the single-channel integration/generation.}
\label{ss3t}
\vspace{-15pt}
\end{center}
}
\end{table}
The results agree well, and exhibit the fact that the generated densities of
the generators the hierarchical type are much closer to the integrand.
The same picture is reproduced for an arbitrary permutation.

For a realistic QCD calculation, the integrated function may be approximated by
a sum over permutations.
Therefore, an efficient generator has to include all possible channels, where
each channel corresponds to a given permutation of the momenta.
In that case, a multi-channeling optimization procedure can be applied, which
is incorporated in {\tt HAAG}.
In order to study the efficiency of this optimization we consider the same
integration as before, but with all channels contributing to the generation and
allowing for optimization.
In this optimization procedure, we discard channels that contribute less than a
certain pre-determined fraction to the set of available channels.
It is expected, of course, that in end the right permutation will be `chosen'
by the optimization.
This is indeed the case and the results are presented in \Table{as3}.
We see that the optimization results to a picture close to the one obtained
with the single channel generation, with some noticeable improvement in the
case of {\tt SARGE}.
\begin{table}[ht]
{\scriptsize
\begin{center}
\begin{tabular}{|c|c|c|c|c|c|c|c|}
\hline
jets & algorithm  & $N_{\mathrm{gen}}$  &  $N_{\mathrm{acc}}$  &  $I$         &   $\Delta I$  & $f$  &          $\varepsilon$(\%)  \\ \hline
\multirow{3}{1cm}{\centering 4}
    & {\tt SARGE}   &  \eee{1}{5}  & 52516   & \eee{.262}{-9} & \eee{.294}{-11} &     $12.6 $         & $1.29$ \\
    & {\tt HAAG}    &  \eee{5}{4}  & 34293   & \eee{.257}{-9} & \eee{.210}{-11} &     $3.36 $         & $4.28$   \\\hline
\multirow{3}{1cm}{\centering 5}
    & {\tt SARGE}   &  \eee{2.5}{5}  & 32315   & \eee{.422}{-10} & \eee{.106}{-11} &     $159  $        & $0.44$ \\
    & {\tt HAAG}    &  \eee{6.5}{4}  & 31063   & \eee{.444}{-10} & \eee{.503}{-12} &     $8.32 $        & $1.17$   \\\hline
\multirow{3}{1cm}{\centering 6}
    & {\tt SARGE}   &  \eee{1}{6}    & 29138   & \eee{.476}{-11} & \eee{.145}{-12} &     $933 $       & $0.45$   \\
    & {\tt HAAG}    &  \eee{1.2}{5}  & 33278   & \eee{.483}{-11} & \eee{.595}{-13} &     $18.2 $      & $1.19$   \\\hline
\end{tabular}
\caption[.]{Results for the all-channel generation with optimization.}
\label{as3}
\vspace{-15pt}
\end{center}
}
\end{table}

As is the case for any multi-channel generator, a computational complexity
problem arises when the number of channels increases.
For instance, in our case we are facing a  number of $\frac{1}{2}(n+1)!$
channels! On the other hand, it is also clear that the channels we are
considering have a large overlap in most of the available phase space.
It is therefore worth to investigate the dependence of the integration
efficiency on the number of channels used.
This is presented in \Table{cha}, where the full antenna
\begin{equation}
 s^2
 \sum_{P(2,\ldots,n+2)}
 [\;(p_1\cdot p_3)(p_3\cdot p_4)(p_4\cdot p_2)(p_2\cdot p_5)
        \ldots(p_{n+2}\cdot p_1)\;]^{-1}
\label{aa}
\end{equation}
is integrated, using a number of channels that has been selected on a random
basis.
We see the rather interesting phenomenon that a decent description can be
achieved with a much smaller number of channels.
\begin{table}[ht]
{\scriptsize
\begin{center}
\begin{tabular}{|c||c|c|c|c|c|c|c|}
\hline
\# channels & $2520$     & $1500$  &  $1000$  &  $500$     &  $200$    &     $50$   &  $10$    \\ \hline
 $f$    & $5.33$     & $5.37$  &  $5.48$  &  $5.72$    &  $6.14$    &    $11.6$  &  $84.7$  \\ \hline
 $N_{\mathrm{acc}}$  & $26630$    & $26521$ &  $26437$ &  $26676$   &  $27009$   &    $27190$ &  $27205$ \\ \hline
 $\varepsilon(\%)$
        & $11.2$     & $13.1$  &  $11.6$  &  $7.1$     &  $7.5$     &    $1.7$   &  $0.28$  \\ \hline
\end{tabular}
\caption[.]{All-channel integration with subsets of channels for generation.}
\label{cha}
\vspace{-15pt}
\end{center}
}
\end{table}
Variations of this technique of using only subsets of channels, for example
choosing another subset after each step of multi-channel optimization, lead to
the same picture.

The complete results of the integration of the full antenna are presented in
\Table{aa30}.
\begin{table}[ht]
{\scriptsize
\begin{center}
\begin{tabular}{|c|c|c|c|c|c|c|c|}
\hline
jets & algorithm  & $N_{\mathrm{gen}}$  &  $N_{\mathrm{acc}}$  &  $I$         &   $\Delta I$  & $\varepsilon$(\%)& $f$    \\ \hline
\multirow{2}{1cm}{\centering 4}
    & {\tt SARGE}&  \eee{1}{5}  & 47483   & \eee{.166}{-7} & \eee{.115}{-9} &     $4.21 $        & $4.8$ \\
    & {\tt HAAG} &  \eee{6}{4}  & 42019   & \eee{.167}{-7} & \eee{.810}{-10} &    $12.01$        & $1.4$   \\\hline
\multirow{2}{1cm}{\centering 5}
    & {\tt SARGE}&  \eee{3}{5}    & 39095   & \eee{.176}{-7} & \eee{.162}{-9} &     $3.27 $        & $25.6$ \\
    & {\tt HAAG} &  \eee{1.2}{5}  & 55234   & \eee{.177}{-7} & \eee{.856}{-10} &    $7.53 $        & $2.7$   \\\hline
\multirow{2}{1cm}{\centering 6}
    & {\tt SARGE}&  \eee{1.5}{6}  & 44529   & \eee{.157}{-7} & \eee{.135}{-9} &     $2.95 $        & $109$ \\
    & {\tt HAAG} &  \eee{1.8}{5}  & 47911   & \eee{.161}{-7} & \eee{.905}{-10} &    $7.15 $        & $5.7$   \\\hline
\multirow{2}{1cm}{\centering 7}
    & {\tt SARGE}&  \eee{1}{7}    & 47766   & \eee{.123}{-7} & \eee{.988}{-10} &     $3.02 $        & $642$ \\
    & {\tt HAAG} &  \eee{3.6}{5}  & 45599   & \eee{.123}{-7} & \eee{.241}{-10} &     $5.11 $        & $13$   \\\hline
\multirow{2}{1cm}{\centering 8}
    & {\tt SARGE}&  \eee{1}{8}    & 53560   & \eee{.784}{-8} & \eee{.554}{-10} &    $3.29$        & $4998$ \\
    & {\tt HAAG} &  \eee{1}{6}   & 49206   & \eee{.789}{-8} & \eee{.496}{-10} &    $6.30$        & $39$   \\\hline
\end{tabular}
\caption[.]{Results for the all-channel integration.}
\label{aa30}
\vspace{-15pt}
\end{center}
}
\end{table}
We see that {\tt HAAG} has a much better $f$ factor than {\tt SARGE}.
On the other hand the $\varepsilon$ exhibits a less dramatic effect.
This is related to the fact that {\tt SARGE} generates a phase space that is
much larger than the one defined by the cut on $s_0$.
In that sense, if the main time consumption in a given computation is spent
over the evaluation of the integrand (matrix element squared), it is more fair
to compare the square of the estimated expected error, normalized by the number
of accepted events $N_{\mathrm{acc}}$.
In that case we see that {\tt HAAG} is still 2-3 times more efficient, and if
we consider a smaller cut, namely $\sqrt{s_0}=10$ GeV, this gain goes up to an
order of magnitude (\Table{aa10}).
\begin{table}[ht]
{\scriptsize
\begin{center}
\begin{tabular}{|c|c|c|c|c|c|c|c|}
\hline
jets & algorithm  & $N_{\mathrm{gen}}$  &  $N_{\mathrm{acc}}$  &  $I$         &   $\Delta I$  & $\varepsilon$(\%)& $f$    \\ \hline
\multirow{2}{1cm}{\centering 4}
    & {\tt SARGE}&  \eee{1}{5}  & 60986   & \eee{.364}{-6} & \eee{.548}{-8} &     $0.631 $        & $22.7$ \\
    & {\tt HAAG} &  \eee{6}{4}  & 46763   & \eee{.366}{-6} & \eee{.235}{-8} &     $4.34$          & $2.47$   \\\hline
\multirow{2}{1cm}{\centering 5}
    & {\tt SARGE}&  \eee{2}{5}    & 43150   & \eee{.619}{-6} & \eee{.165}{-7} &     $0.29 $        & $142$ \\
    & {\tt HAAG} &  \eee{1}{5}    & 56034   & \eee{.643}{-6} & \eee{.465}{-8} &    $1.84 $         & $5.23$   \\\hline
\multirow{2}{1cm}{\centering 6}
    & {\tt SARGE}&    \eee{1}{6}  & 67811   & \eee{.114}{-5} & \eee{.257}{-7} &     $0.28 $        & $502$ \\
    & {\tt HAAG} &  \eee{1.4}{5}  & 51983   & \eee{.111}{-5} & \eee{.883}{-8} &     $2.50 $        & $8.83$   \\\hline
\multirow{2}{1cm}{\centering 7}
    & {\tt SARGE}&  \eee{5}{6}    & 84391   & \eee{.186}{-5} & \eee{.346}{-7} &     $0.176 $        & $1723$ \\
    & {\tt HAAG} &  \eee{2}{5}    & 44015   & \eee{.192}{-5} & \eee{.177}{-7} &     $2.24 $        & $16$   \\\hline
\multirow{2}{1cm}{\centering 8}
    & {\tt SARGE}&  \eee{5}{7}    & 175541   & \eee{.354}{-5} & \eee{.517}{-7} &    $.119$        & $10618$ \\
    & {\tt HAAG} &  \eee{5}{5}    & 58874    & \eee{.350}{-5} & \eee{.289}{-7} &    $1.65$        & $34$   \\\hline
\end{tabular}
\caption[.]%
        {Results for the all-channel integration%
         with $s_0=100\,\mathrm{GeV}^2$.}
\label{aa10}
\vspace{-10pt}
\end{center}
}
\end{table}

\begin{figure}
\begin{center}
\epsfig{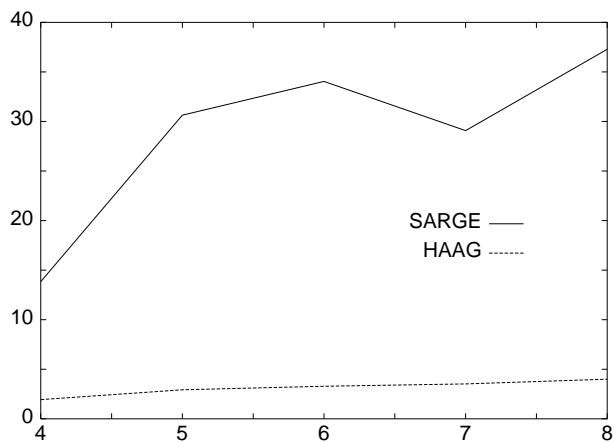}
\caption{
$N_{\mathrm{acc}}f/N_{\mathrm{gen}}$ (a measure of computing time)
as function of the number of produced partons.}
\label{cputime}
\end{center}
\end{figure}

For a realistic calculation of the cross section of a QCD process, one may
assume that the time it takes to perform one evaluation of the integrand is
much larger than the time it takes to generate one accepted event and to
calculate the weight.
This means that the computing time is completely determined by the number of
accepted events $N_{\mathrm{acc}}$.
We introduce
\begin{equation}
   \frac{N_{\mathrm{acc}}f}{N_{\mathrm{gen}}}
\nonumber\end{equation}
as a measure of the computing time.
For a realistic calculation, one has to multiply this number by the evaluation
time of the integrand, and divide by the square of the relative error one wants
to reach.
\Figure{cputime} shows this quantity as function of the number of produced
partons using the data of \Table{aa10}.
According to this graph, a calculation with {\tt SARGE} would take $10$ times
longer than the calculation with {\tt HAAG}.
%

\section{One-loop corrections to electroweak processes}
%
Scattering amplitudes in Quantum Field Theory can be represented by Feynman
diagrams whose number grows extremely rapidly (faster than factorially) in the
number of loops and external legs.
This places severe limits in the calculation of multi-particle scattering
amplitudes.
The last few years, the development of new innovative methods and algorithms
made possible to overcome these limitations.
Based on Schwinger-Dyson recursive equations the complete scattering amplitude
is computed directly without recourse to explicit Feynman diagrams.
This results to a dramatic decrease in the computational cost which now depends
only exponentially, \ie\ $\sim 3^n$, in the number of external particles, $n$.
Automatic computational tools, based on these Schwinger-Dyson recursive
equations, have been developed that are able to describe any process within the
Standard Electroweak theory and QCD~\cite{helac,alpgen,jetI,omega}.

The high precision attainable by the future experiments calls for a reliable
estimate of the higher order corrections to the multi-particle scattering
processes.
This means that the full one-loop contributions along with the higher order
leading QED and electroweak corrections are necessary.
Moreover, taking into account the unstable particle contributions, re-summed
propagator corrections have to be included.

\begin{figure}
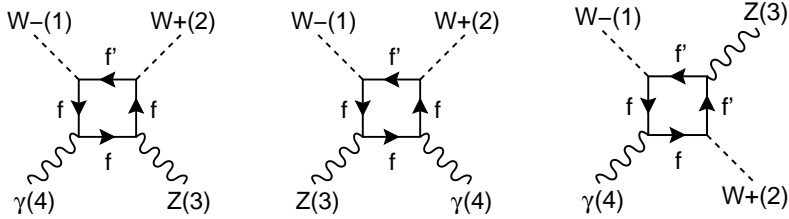

\begin{center}
          \diagram{fig.1}{80}{0}
\quad\quad\diagram{fig.2}{80}{0}
\quad\quad\diagram{fig.3}{82}{0}
\caption{Three of the six 1PI diagrams that contribute to the fermionic
one-loop correction of the $WW\!Z\gamma$ vertex. $f$ represents a down-type
fermion, and $f'$ the corresponding up-type fermion. The other three diagrams
are obtained by exchanging $f\leftrightarrow f'$ and taking the opposite
fermion current.}
\label{diagrams}
\end{center}
\end{figure}

As a first step towards the extension of {\tt HELAC}~\cite{helac} towards the
full one-loop level, re-summed boson propagators and fermion-loop corrections
to boson vertices can be included.
The main reasons to choose his collection of corrections is that it ensures
gauge invariance~\cite{fermionloop}, and that it is fairly straightforward to
implement in an automatic program based on the Schwinger-Dyson method.
If one, for example, wants to analyze processes that do not involve more than
$6$ fermions, the one-loop corrections do not involve diagrams with more than
$4$ external legs.
So the ``most complicated'' diagrams that have to be included are
1PI fermion-one-loop four-point functions (\Figure{diagrams}).
Below, we give a sketch how to evaluate them in a straightforward manner.
The lower-point functions can be evaluated analogously.
%

\subsection{Evaluation of the one-loop four-point function}
Each diagram represents a function of the masses $m_i$ of the fermions, the
momenta $p_i$ of the vector bosons, their polarization vectors $w_i$ and their
couplings $v_i,a_i$ to the fermion current.
We conveniently choose to access them by permutating the input of the general
fermionic one-loop four-point function
\begin{equation}
\Gamma_4
=\int\frac{d^nq}{\mathrm{i}\pi^2}
  \,\mathrm{Tr}[\mathrm{P}_1(q)\mathrm{V}_1\,\mathrm{P}_2(q)\mathrm{V}_2
      \,\mathrm{P}_3(q)\mathrm{V}_3\,\mathrm{P}_4(q)\mathrm{V}_4]
\;\;,
\label{Gamma4}\end{equation}
with \footnote{We define the sum $\sum_{i=1}^{0}x_i$ as a sum of zero terms.}
\begin{equation}
   \mathrm{P}_j(q)
   = \frac{q\!\!\!/+\sum_{i=1}^{j-1}p\!\!\!/_i+m_j}
            {(q+\sum_{i=1}^{j-1}p_i)^2-m_j^2+\mathrm{i}\varepsilon}
   \quad\textrm{and}\quad
   \mathrm{V}_j = w\!\!\!\!/_j(v_j+a_j\gamma_5)
\;\;.
\end{equation}
Possible divergences are treated within the formalism of dimensional
regularization. Ambiguities regarding $\gamma_5$ are avoided if the vectors
$p_i$ and $w_i$ are considered to be strictly $4$-dimensional.
%
%
The trace can be calculated with the help of computer algebra, for example with the program {\tt FORM}~\cite{FORM}, leading to 
\begin{equation}
\Gamma_4 =
   \int\frac{d^nq}{\mathrm{i}\pi^2}
 \,\frac{\mathrm{num}_4(q;p_{1,2,3},m_{1,2,3,4},w_{1,2,3,4}
                               ,v_{1,2,3,4},a_{1,2,3,4})}
        {\mathrm{den}_4(q;p_{1,2,3},m_{1,2,3,4})}
\;\;,
\end{equation}
where the denominator function is defined by
\begin{equation}
   \mathrm{den}_{l}(q;p_{1,2,\ldots,l-1},m_{1,2,\ldots,l})
   = \prod_{j=1}^{l}\bigg[\Big(q + \sum_{i=1}^{j-1}p_i\Big)^2
                          -m_{j}^2+\mathrm{i}\varepsilon\bigg]
\;\;,
\end{equation}
and the numerator $\mathrm{num}_4$ is a fourth-order polynomial in the
components of
$q$.
The integration problem is now reduced to that of the calculation of {\em
tensor\/} integrals of the type
\begin{equation}
 \mathcal{D}_{\nu_1\cdots\nu_r}(p_{1,2,3};m_{1,2,3,4})
 = \int\frac{d^nq}{\mathrm{i}\pi^2}
    \,\frac{q_{\nu_1}\cdots q_{\nu_r}}
        {\mathrm{den}_4(q;p_{1,2,3},m_{1,2,3,4})}
\;\;,
\end{equation}
with $r=0,1,2,3,4$.
The Passarino-Veltman method~\cite{PassarinoVeltman,Denner} uses the Lorentz
covariance of these integrals to express them in terms of {\em coefficient
functions\/} $D$ through
\begin{equation}
   \mathcal{D}_{\nu_1\cdots\nu_r}(p_{1,2,3};m_{1,2,3,4})
  =\!\!\!\sum_{i_1,\ldots,i_r=0}^{3}\!\!\!
     p_{i_1,\nu_1}\cdots p_{i_r,\nu_r}
     D_{i_1i_2\cdots i_r}(p_{1,2,3};m_{1,2,3,4})
\;,
\end{equation}
with $D_{i_1i_2\cdots i_r}=0$ if an odd number of indices are equal to $0$,
and the interpretation
\begin{equation}
   p_{0,\nu_1}p_{0,\nu_2}\cdots p_{0,\nu_{2j}}
  \leftarrow g_{\{\nu_1\nu_2}g_{\nu_3\nu_4}\cdots g_{\nu_{2j-1}\nu_{2j}\}}
\;\;.
\end{equation}
The identification of the $\mathcal{D}$-functions and their expression in terms
of the $D$-functions can also easily be performed by {\tt FORM}.
The output of {\tt FORM}\ will contain symbols representing scalar products of
the external momenta and the polarization vectors, contractions of these with
the Levi-Civita symbol, and the $D$-functions.
This output can easily be turned into a {\tt FORTRAN}-code.
The scalar products and the Levi-Civita symbol are easy to be implemented, and
the $D$-functions can be extracted from the {\tt
LoopTools}-package~\cite{LoopTools}, or one can make the effort to extend the
{\tt FF}-package~\cite{FF}, upon which {\tt LoopTools}\ is based.
%

\subsection{Result}
%
We will present the result of an actual calculation now.  For more results and
more details about the program presented above, we refer
to~\cite{inpreparation}.
There, we will also digress more about the 
well-known problem that the
Passarino-Veltman method to calculate tensor integrals is numerically unstable
for certain values of the external momenta.
It involves the inversion of a 
kinematic matrix, which can become singular, although the one-loop function
is perfectly-well defined. 
On might hope that, in a Monte Carlo calculation of a cross section, the
probability to get too close to these phase-space points is too small to be
concerned about this problem.
We experienced, that this hope may be trusted for cross section calculations
concerning processes at the coming generation of accelerators if the 
computation of the coefficient functions is performed at quadrupole precision.
Restricting the use of quadrupole precision like this, the cpu-time stays
within acceptable limits.
We calculated the total cross section for the process
\begin{displaymath}
 e^- e^+ \to \mu^-\,\bar{\nu}_\mu\,u\,\bar{d}\,\tau^-\,\tau^+
\end{displaymath}
using the following cuts: $ E_l,E_q >5$GeV for lepton and quark energies, a
maximal cosine of $0.985$ between all (initial and final state) charged leptons
and quarks, and $ m_{ll}, m_{qq} > 10$GeV for the invariant masses of charged
leptons and quarks.
We used the renormalization scheme of \cite{paperWIM},
and the following input parameters
%
%
\begin{eqnarray}
  m_{_W} = 80.35\unskip\,\mathrm{GeV} \quad&,&\quad
  m_{_Z} = 91.1867\unskip\,\mathrm{GeV}
\nonumber\\
  \mathrm{Re}[\alpha^{(5)}(m_{_Z}^2)^{-1}] = 128.89  \quad&,&\quad
  {\alpha(0)}^{-1} = 137.03599976
\\
  G_F= 1.16639&\times& 10 ^{-5}\unskip\,\mathrm{GeV} \;\;.
\nonumber\end{eqnarray}
As far as the tree-order cross section is concerned we use the widely used
Fixed Width  scheme, where a fixed $W$-boson width is implemented in all
$W$-boson propagators and where the $G_F$-scheme is applied for evaluating the
weak parameters.
We recall that the latter is defined by using $m_{_W}$, $m_{_Z}$ and $G_F$ as
input parameters, together with the two relations
\begin{displaymath}
  s_{_W}^2 = 1 - \frac{m_{_W}^2}{m_{_Z}^2} \quad,\quad
  \alpha = \frac{\sqrt{2}}{\pi}\,G_F\,m_{_W}^2 s_{_W}^2\;\;.
\end{displaymath}
For the $W$ and $Z$ widths we use
\begin{displaymath}
 \Gamma_W=2.042\mathrm{GeV} \quad,\quad \Gamma_Z=2.49\mathrm{GeV} \;\;.
\end{displaymath}
The results for $E=500$GeV are $\sigma_0/ab=54.96(26)$
$\sigma_1/ab=57.31(28)$, and the K-factor is $K/100=4.28(2)$.
They show the expected contribution at the percent level of the higher order FL
corrections to the total cross section.
For comparison, the value of $\sigma_0/ab$ computed with {\tt
O'Mega/WHIZARD}~\cite{omega,whizard} is given by $55.07(19)$.

\section{Conclusions}
%
We presented the algorithm {\tt HAAG}, which uses the ``hierarchical'' strategy
for phase space generation in order to efficiently map the antenna momentum
structures, typically occurring in  QCD amplitudes. 
It exhibits an improved efficiency compared to {\tt SARGE} for multi-parton
calculations, and it is more powerful in describing densities where a partial
symmetrization over the permutation space is considered.
Also, {\tt HAAG} makes no fundamental distinction among massless and massive
particles, so it can be used for an arbitrary multi-partonic process.

Furthermore we implemented one-loop corrections following the Fermion-Loop
scheme in the program {\tt HELAC} for automatic amplitude calculation and 
presented a result of a cross section calculation.

\section*{Acknowledgments}
%
The authors would like to thank C.\ Schwinn for providing the result of the
cross section calculation with {\tt O'Mega/WHIZARD}.
The research has been financially supported by the European Union under
contract number HPRN-CT-2000-00149, and through a Marie Curie Fellowship under
contract number HPMD-CT-2001-00105
%

\def\EPJ#1#2#3{Eur.\ Phys.\ J.\ {\bf #1} (#2) #3}
\def\CPC#1#2#3{Comp.\ Phys.\ Comm.\ {\bf #1} (#2) #3}
\def\PR#1#2#3{Phys.\ Rev.\ {\bf{#1}} (#2) #3}
\def\PRL#1#2#3{Phys.\ Rev.\ Lett.\ {\bf{#1}} (#2) #3}
\def\PL#1#2#3{Phys.\ Lett.\ {\bf{#1}} (#2) #3}
\def\PRep#1#2#3{Phys.\ Rep.\ {\bf{#1}} (#2) #3}
\def\NP#1#2#3{Nucl.\ Phys.\ {\bf{#1}} (#2) #3}
\def\ZP#1#2#3{Z.\ f.\ Phys.\ {\bf{#1}} (#2) #3}
\def\IJMP#1#2#3{Int.\ J.\ Mod.\ Phys.\ {\bf{#1}} (#2) #3}
\def\MPL#1#2#3{Mod.\ Phys.\ Lett.\ {\bf{#1}} (#2) #3}
\def\ibid#1#2#3{{\it ibid} {\bf{#1}} (#2) #3}
\def\JHEP#1#2#3{JHEP {\bf{#1}} (#2) #3}
\def\FP#1#2#3{Fortsch.\ Phys.\ {\bf #1} (#2) #3}

\end{document}